\documentclass[a4paper]{jpconf}
\usepackage{graphicx}
\begin{document}
\newcommand{\newc}{\newcommand}
\newc{\ra}{\rightarrow}
\newc{\lra}{\leftrightarrow}
\newc{\beq}{\begin{equation}}
\newc{\eeq}{\end{equation}}
\newc{\barr}{\begin{eqnarray}}
\newc{\earr}{\end{eqnarray}}
\newcommand{\Od}{{\cal O}}
\newcommand{\gsim}   {\mathrel{\mathop{\kern 0pt \rlap
  {\raise.2ex\hbox{$>$}}}
  \lower.9ex\hbox{\kern-.190em $\sim$}}}
  \def\rpm{R_p \hspace{-0.8em}/\;\:}
\title{Review of Low Energy Neutrinos}

\author{J.D. Vergados}

\address{University of Ioannina, Ioannina GR 451 10, Greece}

\ead{vergados@cc.uoi.gr}

\begin{abstract}
Some issues regarding low energy neutrinos are reviewed. We focus on three aspects i)We show that by employing  very low energy (a few keV) electron neutrinos, neutrino disappearance  oscillations can be investigated by detecting recoiling electrons with low threshold spherical gaseous TPC's. In such an  experiment, which is sensitive to the small
 mixing angle $\theta_{13}$, the novel feature is that the oscillation length is so small 
that the full oscillation
takes place inside the detector. Thus one can determine accurately all the oscillation parameters and,
in particular,  measure or set
 a good limit on $\theta_{13}$. ii) Low threshold gaseous TPC detectors can also be used in detecting nuclear recoils by
exploiting the neutral current interaction. Thus these robust and stable detectors can be employed in supernova neutrino detection. iii) The lepton violating neutrinoless
 double decay is investigated focusing on how the absolute neutrino mass can be extracted from the data.
\end{abstract}
\section{Introduction.}

The discovery of neutrino oscillations can be considered as one of the greatest triumphs of modern physics.
It began with atmospheric neutrino oscillations \cite{SUPERKAMIOKANDE} interpreted as
 $\nu_{\mu} \rightarrow \nu_{\tau}$ oscillations, as well as
 $\nu_e$ disappearance in solar neutrinos \cite{SOLAROSC}. These
 results have been recently confirmed by the KamLAND experiment \cite{KAMLAND},
 which exhibits evidence for reactor antineutrino disappearance.
  As a result of these experiments we have a pretty good idea of the neutrino
mixing matrix and the two independent quantities $\Delta m^2$, e.g $|m_2^2-m^2_1|$ and $|m^2_3-m^2_2|$.
 Fortunately these
two  $\Delta m^2$ values are vastly different \cite{MALTONI04}, see Eq. (\ref{matdat}) below.
 This means that the relevant $L/E$ parameters are very different. Thus for a given energy the experimental results can approximately be described as two generation oscillations. For an accurate description  of the data, however, a three generation analysis is necessary. 

 In all of these analyses the
oscillation length is much larger than the size of the detector. So one is able to see the effect, if the detector is
placed in the right distance from the source. It is, however, possible to design an experiment with an oscillation
length of the order of the size of the detector. 
 This is achieved, if one considers a neutrino source with as low as practical average neutrino energy,
such as a triton source with a maximum energy of $18.6$ keV. Thus the average oscillation length is $6.5$m, which is
smaller than the radius of $10$m of a spherical gaseous TPC detector \cite{NOSTOSA}. Such low energy events can be detected by measuring  recoiling electrons with a low threshold spherical TPC detector. 

In a typical supernova an energy of about $10^{53}$ ergs is released in the form of neutrinos
\cite{BEACFARVOG},\cite{SUPERNOVA}. These neutrinos
are emitted within an interval of about $10$ s after the explosion and they travel to Earth undistorted, except that,
on their way to Earth, they may oscillate into other flavors. 
Thus for traditional detectors
 relying on the charged current
 interactions the precise event rate may depend critically on the specific properties of the neutrinos. The
time integrated spectra in the case of charged current detectors, like the SNO experiment, 
depend on the neutrino oscillations \cite{TKBT03}. 
Recently it has become feasible to detect neutrinos by exploiting the neutral current interaction \cite{DKLEIN} and measuring
 the recoiling nucleus. One employs   gaseous TPC detectors with low threshold energies. 
A description of the NOSTOS project and details of the spherical TPC detector with sub-keV threshold 
 are given in \cite{NOSTOSA},\cite{NOSTOSB}.
 The whole system looks stable, robust and easy to maintain.
The neutral current interaction, through its vector component, can lead
 to coherence, i.e. an additive contribution of all neutrons in the nucleus (the vector  contribution of the
 protons is tiny).
 
  Finally, in spite of the great process been made in understanding neutrinos mainly via neutrino oscillations,
  some fundamental issues remain unsettled. First are the neutrinos Dirac or Majorana particles? What is the absolute 
  scale of the neutrino masses? The first question can practically be answered only by neutrinoless double beta
  decay, see, e.g., earlier reports \cite{VERGADOS} and references therein . Furthermore neutrinoless double beta decay offers the best hope for answering the second question down to
  a few meV.
  \section {Coherent neutrino nucleus scattering}
The standard neutral current  left handed weak interaction can be cast in the form:
\begin{equation}
{\cal L_q}=-\frac{G_F}{\sqrt{2}}
\left[ \bar\nu_\alpha \gamma^\mu (1-\gamma^5)\nu_\alpha \right]
\left[ \bar q\gamma_\mu (g_V(q)-g_A(q)\gamma^5) q \right] 
\label{weak1}
\end{equation}
(diagonal in flavor space).
At the nucleon level we get:
\begin{equation}
{\cal L_q}=-\frac{G_F}{\sqrt{2}}
\left[ \bar\nu_\alpha \gamma^\mu (1-\gamma^5)\nu_\alpha \right]
\left[ \bar N\gamma_\mu (g_V(N)-g_A(N)\gamma^5) N \right] 
\label{weak3}
\end{equation}
 with
\begin{equation}
g_V(p)=\frac{1}{2}-2 \sin^2{\theta_W}\simeq 0.04~~,~~g_A(p)=1.27 \frac{1}{2}~;~
g_V(n)=-\frac{1}{2}~~,~~g_A(n)=-\frac{1.27}{2}
\label{weak4}
\end{equation}
Beyond the standard level one has further interactions which need not be diagonal in flavor 
space \cite{VERGIOM}. We are not. however, going to discuss such issues here.
The cross section for elastic neutrino nucleon scattering has extensively been studied 
\cite{BEACFARVOG},\cite{VogEng}.

The energy of the recoiling particle can be written in dimensionless form
as follows:
\beq
y=\frac{2\cos^2{\theta}}{(1+1/x_{\nu})^2-\cos^2{\theta}}~~,~~
y=\frac{T_{recoil}}{m_{recoil}},x_{\nu}=\frac{E_{\nu}}{m_{recoil}}
\label{recoilen}
\eeq 
 The maximum  energy occurs when $\theta=0$, $y_{max}=\frac{2}{(1+1/x_{\nu})^2-1}$,
in agreement with Eq. (2.5) of earlier work. \cite{BEACFARVOG}.
  One can invert Eq. \ref{recoilen} and get the  neutrino energy associated with a given recoil energy and
scattering angle.
From the above expressions we see that the vector current contribution, which may lead to coherence, is negligible
in the case of the protons. Thus the coherent contribution  may come from the neutrons and is expected to be
proportional to the square of the neutron number.
  The neutrino-nucleus coherent cross section takes the form:
 
   \begin{eqnarray}
 \left(\frac{d\sigma}{dT_A}\right)_{weak}&=&\frac{G^2_F Am_N}{2 \pi}~(N^2/4)F_{coh}(A,T_A,E_{\nu}),
\nonumber\\
& &F_{coh}(A,T_A,E_{\nu})=F(T_A) {\huge [}
  \left (1+\frac{A-1}{A}\frac{T_A}{E_{\nu}} \right )
+(1-\frac{T_A}{E_{\nu}})^2
\nonumber\\
& &\left (1-\frac{A-1}{A}\frac{T_A}{m_N}\frac{1}{E_{\nu}/T_A-1} \right )
-\frac{Am_NT_A}{E^2_{\nu}} ]
 \label{elaswAV}
  \end{eqnarray}
where $F(T_A)$ is the nuclear form factor.
\section{Supernova Neutrinos}
\label{sec.supernova}
\begin{figure}[h]
\begin{minipage}{18pc}
 \rotatebox{90}{\hspace{1.0cm} {$F_{coh}$}}
\includegraphics[width=18pc]{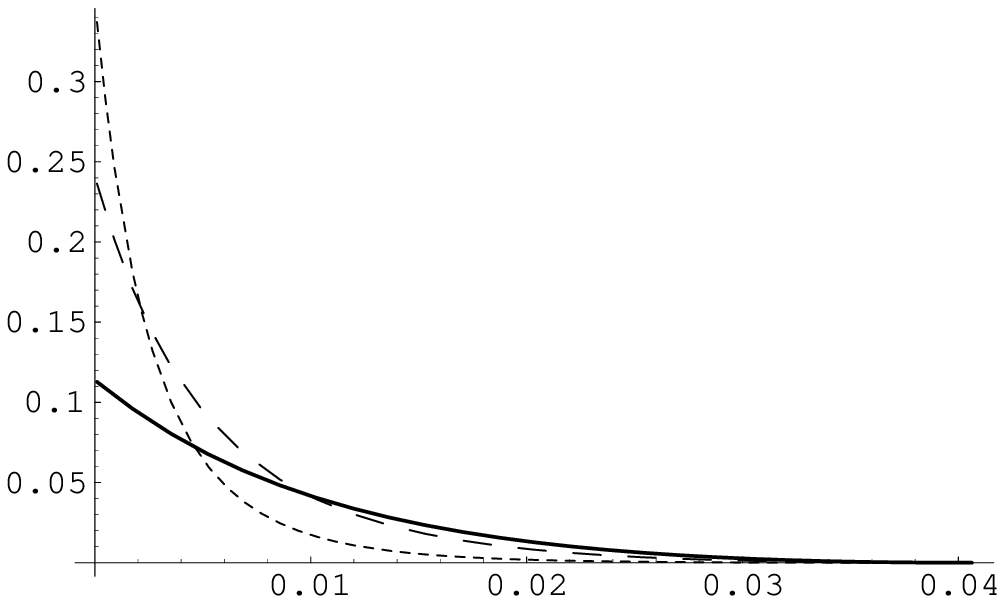}
\hspace*{6.0cm}\tiny{$T_A \rightarrow$ MeV}
\caption{\label{fig:difr131a}The differential event rate as a function of the recoil energy $T_A$, in arbitrary units, for
Xe without quenching.}
\end{minipage}\hspace{2pc}%
\begin{minipage}{18pc}
 \rotatebox{90}{\hspace{1.0cm} {$F_{coh}$}}
\includegraphics[width=18pc]{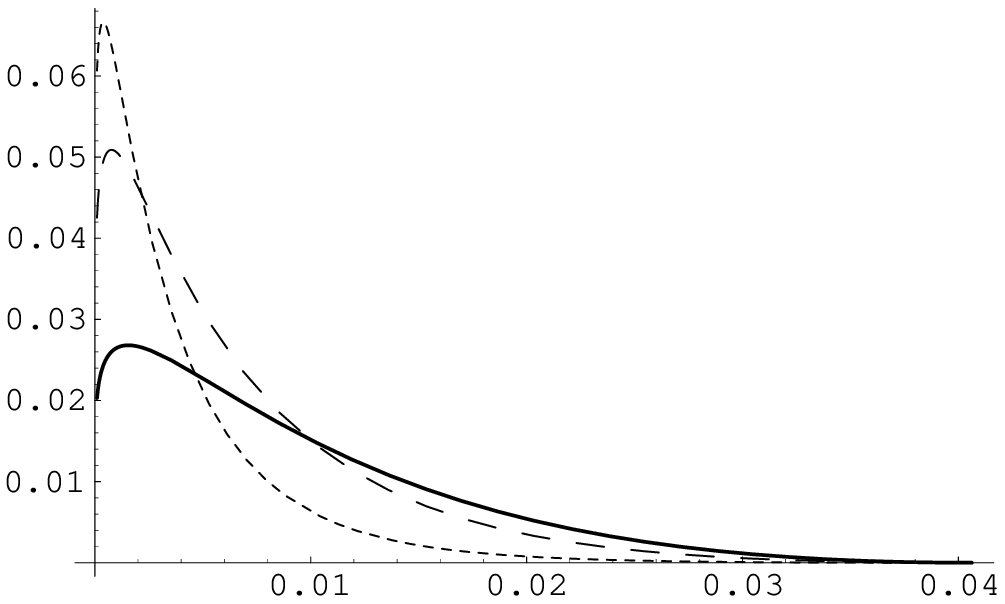}
\hspace*{6.0cm}\tiny{$T_A \rightarrow$ MeV}
\caption{\label{fig:difr131b}The same as in \ref{fig:difr131a} with quenching included.}
\end{minipage} 
\end{figure}

The number of neutrino events for a given detector depends on the neutrino spectrum and the distance of the
source. We will consider a typical case of a source which is about $10$ kpc, l.e. $D=3.1 \times 10^{22}$ cm ( of the order of the radius of the galaxy) with 
an energy output of $3 \times 10^{53}$ ergs with a duration of about $10$ s.  Furthermore we will assume for simplicity that each neutrino flavor is characterized by  a 
 Fermi-Dirac like distribution times its characteristic cross section and we will not consider here the more
realistic distributions, which have recently become available \cite{NUSPECTRA}. 
This is adequate for our purposes. Thus: 
\beq
\frac{dN}{dE_{\nu}}=\sigma(E_{\nu})\frac{E^2_{\nu}}{1+exp(E_{\nu}/T)}=\frac{\Lambda}{JT}\frac{x^4}{1+e^x}~~,
~~x=\frac{E_{\nu}}{T}
\label{nudistr}
\eeq
with
$J=\frac{31\pi^6}{252}$, $\Lambda$  a constant and 
$T$ the temperature of the emitted neutrino flavor. 
Each flavor is characterized by its
own temperature as follows:
$$T=8 \mbox { MeV for } \nu_{\mu},\nu_{\tau},\tilde{\nu}_{\mu}, \tilde{\nu}_{\tau}
\mbox{ and } T=5 ~(3.5)\mbox{ MeV for } \tilde{\nu}_e ~(\nu_e)$$
The constant $\Lambda$ is determined by the requirement that the distribution yields the total energy of each
neutrino species.
$$U_{\nu}=\frac{\Lambda T}{J}\int_0^{\infty } dx \frac{x^5}{1+e^x}\Rightarrow \Lambda=\frac{U_{\nu}}{T}$$
We will further assume that  $U_{\nu}=0.5 \times 10^{53}$ ergs
per neutrino flavor. Thus one finds:
$$\Lambda=0.89\times 10^{58}~(\nu_e),~~0.63\times 10^{58}~(\tilde{\nu}_e)~,0.39\times 10^{58}
\mbox{ (all other flavors)}$$
  The differential event rate (with respect to the recoil energy) is proportional to the quantity:
\beq
\frac{dR}{dT_A}=\frac{\lambda (T)}{J}\int_0^{\infty } dx
F_{coh}(A,T_A,xT) \frac{x^4}{1+e^x}
\label{dRdT}
\eeq
with $\lambda(T)=(0.89,0.63,0.39)$  for $\nu_e,\tilde{\nu}_e$ and all other flavors respectively. 
This is shown  in Figs. \ref{fig:difr131a} and \ref{fig:difr131b}.
Summing over all the neutrino species we can write \cite{NOSTOSB}:
\beq
\mbox{No of events}=C_{\nu} \frac{K(A,(T_A)_{th})}{K(40,(T_A)_{th})}Qu(A)
\label{sumevents}
\eeq
with
\beq
C_{\nu}=153  \left ( \frac{N}{22} \right )^2 \frac{U_{\nu}}{0.5\times 10^{53}ergs}
\left ( \frac{10kpc}{D}\right )^2 \frac{P}{10Atm}
\left[ \frac{R}{4m}\right]^3 \frac{300}{T_0}
\label{C2}
\eeq

\begin{figure}[h]
\begin{minipage}{18pc}
 \rotatebox{90}{\hspace{-0.0cm} {\tiny $K(A,(T_A)_{th})\rightarrow $}}
\includegraphics[width=18pc]{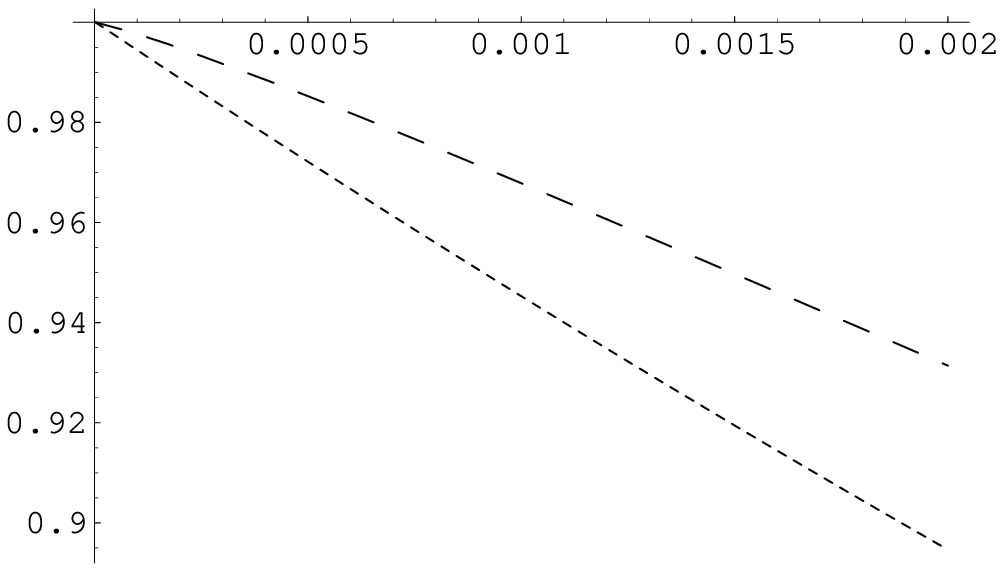}
\hspace*{6.0cm}\tiny{$T_A \rightarrow$ MeV}
\caption{\label{fig:Kqua}The function $K(A,(T_A)_{th})$ versus $(T_A)_{th}$ for the target Ar. The short and long dash correspond to no quenching and quenching factor respectively.
For a threshold energy of $100$ eV the rates are quenched by factors of $3$ (see Eq. \ref{quench2}).}
\end{minipage}\hspace{2pc}%
\begin{minipage}{18pc}
 \rotatebox{90}{\hspace{1.0cm} {$F_{coh}$}}
\includegraphics[width=18pc]{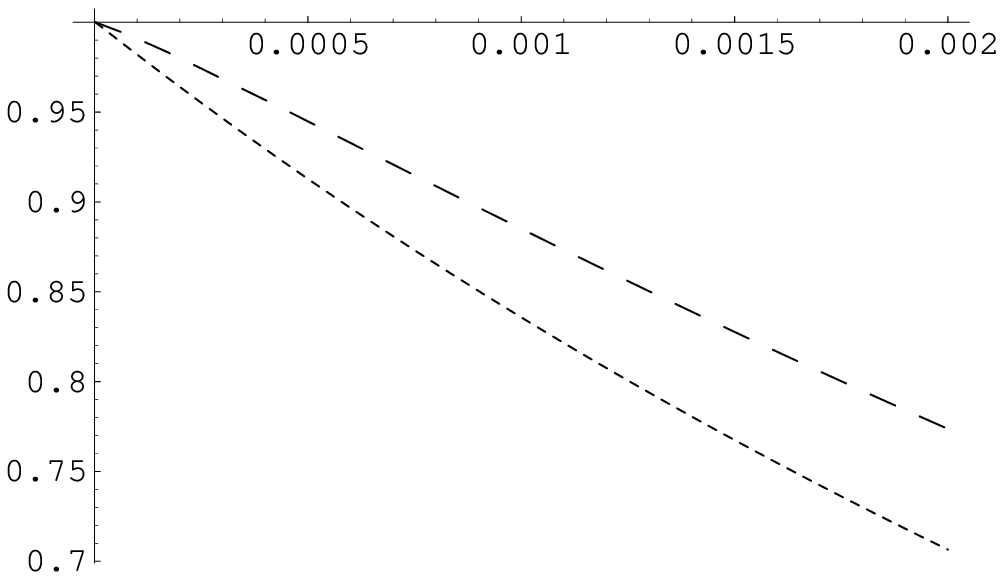}
\hspace*{6.0cm}\tiny{$T_A \rightarrow$ MeV}
\caption{\label{fig:Kqub}The same as in \ref{fig:Kqua} in the Xe. The effect of quenching now is $3.5$ }
\end{minipage} 
\end{figure}

 $K(A,(T_A)_{th})$
 is  the rate at a given threshold energy divided by that at zero threshold. It depends
 on the threshold
energy, the assumed quenching factor and the nuclear mass number. It is unity at $(T_A)_{th})=0$.
 From the above equation we find
that, ignoring quenching, the following expected number of events:
\beq
1.25,~31.6,~153,~614,~1880\mbox{ for He, Ne, Ar, Kr and Xe}
\label{allrates}
\eeq
respectively. For other possible targets the rates can be found by the above formulas or interpolation.\\
 The quantity $Qu(A)$ is the quenching factor  \cite{SIMON03}-\cite{LIDHART}, assuming a  threshold energy
  $(T_A)_{th}=100$eV.
The parameter $Qu(A)$ takes the values:
\beq
0.49,~0.38,~0.35,~0.31,~0.29\mbox{ for He, Ne, Ar, Kr and Xe}
\label{quench2}
\eeq
respectively. The effect of quenching is larger in the case of  heavy targets, since the average energy of
the recoiling nucleus is smaller.
  The effect of quenching is exhibited in Figs \ref{fig:Kqua}  and \ref{fig:Kqub} for the two interesting targets 
Ar and Xe.
 We should mention that it is of paramount importance to experimentally measure the quenching factor. The
 above estimates were based on the assumption of a pure gas. 
 Such an effect will lead to an increase  in the quenching factor and needs be measured.\\
  Anyway the number of expected events including quenching and $E_{th}$=0.1 keV becomes:
\beq
0.61,~  12.0,~  53.5,~  190,~  545  \mbox{ for He, Ne, Ar, Kr and Xe}       
\eeq
The inclusion of the form factor is important only in the case of Xe, in which case the above number of events
 becomes $415$. 
 \section{Neutrino oscillations}
 The neutrino mixing can be
parametrized as follows:
\begin{equation}
\rm{U}=
\left(
\begin{array}{ccc}
1 & 0 & 0\\
0 & c_{23} & s_{23} \\
0 & -s_{23} & c_{23}
\end{array} \right)
\left(
\begin{array}{ccc}
\nonumber
c_{13} & 0 & s_{13}~e^{i\delta}\\
0 & 1 & 0 \\
-s_{13}~e^{i\delta} & 0 & c_{13}
\end{array} \right)
\left(
\begin{array}{ccc}
c_{12} & s_{12} & 0\\
-s_{12} & c_{12} & 0 \\
0 & 0 & 1
\end{array} \right)
\nonumber
\end{equation}
where $s_{ij}=\sin{\theta_{ij}}$  and $c_{ij}=\cos{\theta_{ij}}$.

\begin{figure}[h]
\begin{minipage}{18pc}
\includegraphics[width=18pc]{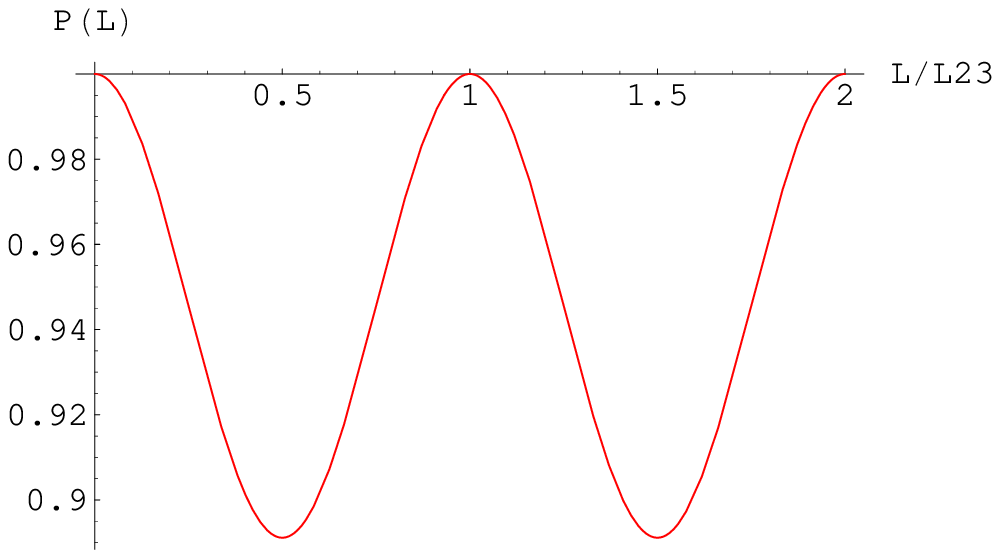}
\caption{\label{triton}The small oscillation length $\nu_e$ disappearance expected to be seen in a TPC
low energy electron detector.}
\end{minipage}\hspace{2pc}%
\begin{minipage}{18pc}
\includegraphics[width=18pc]{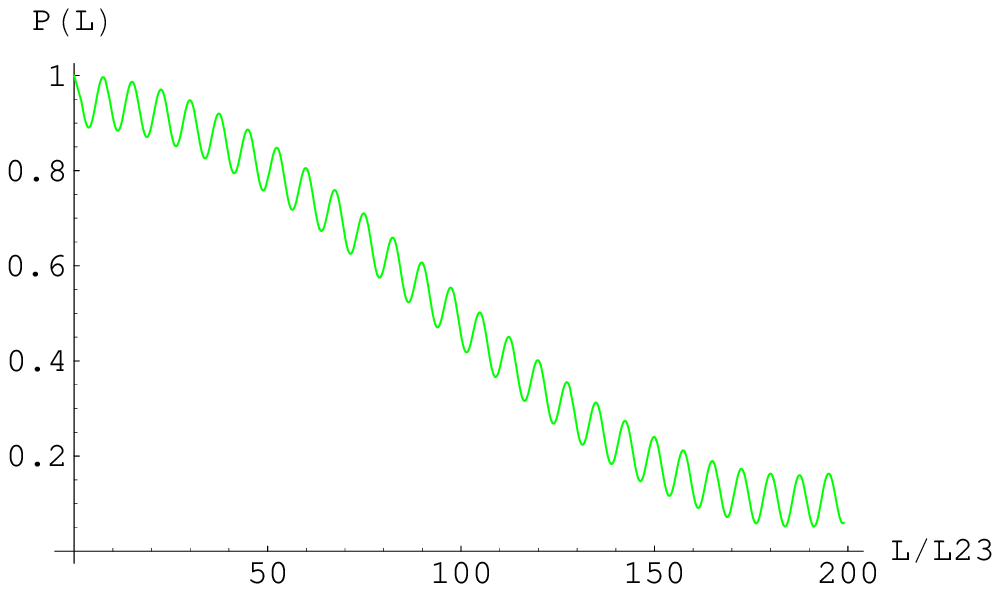}
\caption{\label{combined} On top of the large oscillation length seen in reactor experiments we show the small oscillation
length due to $\theta_{13}$.}
\end{minipage}
\end{figure}

The neutrino oscillation data can be summarized as follows \cite{MALTONI04}:
\beq
\left(
\begin{array}{cccc}
\mbox {parameter}& \mbox{best fit} & 2 \sigma &3 \sigma \\
&&&\\
\Delta m^2_{31}(10^{-3} \mbox {eV}^2) & 1.3 & 1.7-2.9&1.4-33\\
\Delta m^2_{21}(10^{-5} \mbox {eV}^2 )& 8.1 & 7.3-8.7&7.2-9.1 \\
\sin^2{\theta_{12}} & 0.3 & 0.25-0.34&0.23-0.38\\
\sin^2{\theta_{23}} & 0.5 & 0.38-0.64&0.38-0.68\\
\sin^2{\theta_{13}} & 0.00 &\preceq  0.028&\preceq  0.047
\end{array} \right)
\label{matdat}
\eeq
 In a three generation model the electron neutrino disappearance probability is given by:
\barr
    P(\nu_e \rightarrow \nu_e)&=1& { -\cos^2{\theta_{13}} \sin^2{2 \theta_{12}} \sin^2 {\Delta_{21}}}
    \nonumber\\
    &&{-\cos^2{\theta_{12}} \sin^2 {2 \theta_{13}} \sin^2{ (\Delta_{32}-\Delta_{21})}} { -
      \sin^2{\theta_{12}} \sin^2{2 \theta_{13}} \sin^2{\Delta_{32}}}
\earr
     with
 \beq
{\Delta_{21}=\frac{\Delta m^2_{21} L}{2 E_{\nu}}~,~\Delta_{32}=\frac{\Delta m^2_{31} L}{2 E_{\nu}}}
\eeq
In the presence of neutrino mixing both oscillation lengths contribute to the electron neutrino
disappearance.
    For $|\Delta_{21}|\ll |\Delta_{32}|$ and $\theta_{13}\ll 1 $ we get
\beq
     P(\nu_e \rightarrow \nu_e)=1-\sin^2{2 \theta_{12}} \sin^2{\frac{\Delta m^2_{21} L}{2 E_{\nu}}}-
     \sin^2{2 \theta_{13}} \sin{\frac{\Delta m^2_{32} L}{2 E_{\nu}}} 
     \eeq
  \beq{ P(\nu_e \rightarrow \nu_e)\simeq 1} { -\sin^2{2 \theta_{12}} \sin^2{\pi \frac{L} {50 L_{32}}}} {-
     \sin^2{2 \theta_{13}} \sin{\pi \frac{L}{L_{32}}} }
\eeq
     with
     $${ L_{32}=\frac{2 E_{\nu}}{\pi \Delta^2_{32}}}= \mbox{ { small oscillation length}}~,~
L_{12}\approx50L_{23}= \mbox{  large oscillation length}$$
These are shown in Figs \ref{triton}-\ref{combined}.

\section{Neutrino mass Limits from astrophysics and triton decay}

The neutrino oscillation data alone cannot determine the absolute neutrino mass scale and the sign of $\Delta m^2_{32}$. 
We thus distinguish the following cases
\begin{itemize}
\item { Normal Hierarchy}:
$${ \Delta m^2_{SUN}=m^2_2-m^2_1}~,~{ \Delta m^2_{ATM}=m^2_3-m^2_1}$$
$${ m_1}~,~m_2=\sqrt{{ \Delta m^2_{SUN}}+{ m_1^2}}~,~m_3=\sqrt{{\Delta m^2_{ATM}}+{ m_1^2}}$$
\item {Inverted Hierarchy}:
$${\Delta m^2_{SUN}=m^2_2-m^2_1}~,~{\Delta m^2_{ATM}=m^2_2-m^2_3}$$
$${m_3}~,~m_2=\sqrt{{ \Delta m^2_{ATM}}+{ m_3^2}}~~,
~~~m_1=\sqrt{{\Delta m^2_{ATM}-\Delta m^2_{SUN}}+{ m_3^2}}$$
\item The degenerate scenario.
$$m_1=m_2=m_3>>\sqrt{|\Delta m^2_{23}|}$$
\end{itemize}
These problems can be tackled from other experiments as  follows:
\begin{enumerate}
\item The astrophysics limit as follows: 
\begin{itemize}
\item { Normal Hierarchy}:
$${m_1}+\sqrt{{ \Delta m^2_{SUN}}+{ m_1^2}}+\sqrt{{\Delta m^2_{ATM}}+{ m_1^2}}\leq { m_{astro}}$$
\item { Inverted Hierarchy}:
$${ m_3}+\sqrt{{ \Delta m^2_{ATM}}+{ m_3^2}}+
\sqrt{{ \Delta m^2_{ATM}-\Delta m^2_{SUN}}+{m_3^2}}\leq { m_{astro}}$$
\end{itemize}
unfortunately at present this limit is not very stringent \cite{COSMO06} $ m_{astro}<0.71$eV
\item The triton decay limit  
\begin{itemize}
\item { Normal Hierarchy}:
$$c_{12}^2 c^2_{13}{ m^2_1}+s_{12}^2 c^2_{13}({ \Delta m^2_{SUN}}+{ m_1^2})+s^2_{13}({\Delta m^2_{ATM}}+{ m_1^2})\leq { m^2_{decay}}$$
\item {Inverted Hierarchy}:
$$s^2_{13}{ m^2_3}+s_{12}^2c^2_{13}({ \Delta m^2_{ATM}}+{ m_3^2})++c^2_{12}c^2_{13}({ \Delta m^2_{ATM}-\Delta m^2_{SUN}}+{ m_3^2})
\leq { m^2_{decay}}$$
\end{itemize}
This limit is also not very stringent \cite{DECAY02} $m_{decay}<$2.2eV
\end{enumerate}
The only process which at present offers the best chance of reaching limits comparable to the neutrino oscillation
data is neutrinoless double beta decay.
\section{Neutrinoless Double Beta Decay.}
Double beta decay of a nucleus A(N,Z) occurs when the ordinary $\beta$ decay to the nucleus $A(N,Z\pm 1)$ is forbidden due to
energy conservation or angular momentum mismatch, while the decay to one of the nuclei $A(N,Z\pm 2)$ is allowed.
Ignoring the non exotic decays involving neutrinos, the following decays are possible:
$$N(A,Z) \rightarrow N(A,Z+2)+e^- +e^- ~ (0\nu ~\beta \beta ~\mbox {-decay})$$
(the corresponding two neutrino decay has already been observed in many systems).
Furthermore, omitting the non exotic processes accompanied by neutrinos, the following processes
are possible:
$$N(A,Z) \rightarrow N(A,Z-2)+e^+ +e^+ 
  \mbox {(double ~positron~ emission)}$$
$$ e^-_b + N(A,Z)~\rightarrow N(A,Z-2)+e^+ \mbox
   {(electron ~positron ~conversion)}$$
$$e^- _b + e^- _b +N(A,Z) \rightarrow \mbox {N(A,Z-2)+2 X-rays
   (Double electron capture)}$$
   We will limit our discussion on the first of these processes, which is the most important experimentally.
   We will adopt the most popular view and assume that the process proceeds via intermediate neutrinos.
   The relevant half life time is given by:
   \beq
   [T_{1/2}^{0\nu}]^{-1} = G^{0\nu}_{01} \left |\frac{<m_{\nu}>}{m_e} \Omega_{\nu} \right |^2
   \eeq
   where $G^{0\nu}_{01}$ is well understood kinematical factor, $\Omega_{\nu}$ is the nuclear matrix
element and $<m_{\nu}>$ is the average neutrino mass given by: 
\beq
\left < m_{\nu} \right >=c_{12}^2 c_{13}^2 m_1 +s^2_{12}c_{13}^2 e^{i\alpha} m_2+s_{13}^2  e^{i\beta} m_3
\eeq
with $c_{ij}=\cos{\theta_{ij}}$, $s_{ij}=\sin{\theta_{ij}}$,  $m_i$, $i=1,2~3$ are the neutrino masses and $\alpha$ and $\beta$ the two relevant Majorana phases. 

Once the nuclear matrix elements are known $<m_{\nu}>$ can be extracted from the data. From this the lightest neutrino mass
can be inferred, if  neutrino oscillation data are incorporated. This analysis has already been
 done \cite{PETCOV06}, \cite{VALLE06} 
and we are not going to elaborate on it here. The main conclusions are:
If the degenerate scenario holds double beta decay observation is 
within the goals of the current experiments. If the inverted hierarchy holds, there exists a lower bound on 
the value of $<m_{\nu}>$ which is within the reach of the currently planned future experiments. If, however, the 
normal hierarchy scenario holds there is no such lower bound and the road towards measuring $<m_{\nu}>$ may be very long. 
The above conclusions depend, however, on the assumption that the neutrino mass mechanism dominates in $0\nu$ $\beta\beta$ decay.

There exist many other mechanisms which may contribute to double beta decay. The most prominent are intermediate
heavy neutrinos
and R-parity, and consequently lepton violating, supersymmetry \cite{VERGADOS}.

We will extend our formalism to consider the case of right handed currents: The mixing matrix is a $6\times 6$ and takes the form:
\beq
 U=\left ( \begin{array}{c}\nu^0_L\\
 \nu^{0c}_{L} 
 \end{array} \right )
\left ( \begin{array}{cc}
U^{11}~~~&U^{12}\\
U^{21}~~~&U^{22}
 \end{array} \right )
\left ( \begin{array}{c}\nu_L \\
N_L\end{array} \right )
\eeq

where
$$\nu^0_L=(\nu_{eL},\nu_{\mu L},\nu_{\tau L})~;~
\nu^{0c}_L=(\nu^c_{eL},\nu^c_{\mu L},\nu^c_{\tau L})~\Longleftrightarrow  \mbox{ right~handed~neutrino}$$
$$\nu_L=(\nu_{1L},\nu_{2L},\nu_{3L}) \mbox { light}~,~N_L=(N_{1L},N_{2L},N_{3L})\mbox { heavy}$$

If the right handed neutrino does not exist:
{ $$U=U^{11}=U_{MNS}$$
}Quite generally the half-life takes the form \cite{VERGADOS}:
	\begin{eqnarray}
\lefteqn{
[T_{1/2}^{0\nu}]^{-1} = G^{0\nu}_{01}  
\left\{ |X_{L}|^2 + |X_{R}|^2 - 
{\tilde C}_1^\prime X_{L} X_{R}+...
\right.  } & & \nonumber \\ 
&&+ {\tilde C}_2 |\lambda| X_{L} cos \psi_1 
 + {\tilde C}_3 |\eta| X_{L} cos \psi_2 
+ {\tilde C}_4 |\lambda|^2 + {\tilde C}_5 |\eta|^2 
\nonumber \\
&& \left. + {\tilde C}_6 
|\lambda||\eta| cos (\psi_1 -\psi_2) + Re ({\tilde C}_2 
\lambda X_{R} + {\tilde C}_3 \eta  X_{R}) \right\},
\end{eqnarray} 
Where the left handed contribution is:
	\begin{eqnarray}
X_{L}^{} = \eta_{\nu} \Omega_{\nu}+\eta^L_N \Omega_N+\eta_{SUSY} \Omega_{SUSY}
\end{eqnarray} 
with $\Omega_{\nu}~,~\Omega_N~,~\Omega_{SUSY}$ the nuclear matrix elements, associated with light neutrinos, heavy 
neutrinos and SUSY contributions respectively, while  
$\eta_{\nu}~,~\eta^L_N~,~\eta_{SUSY}$ lepton violating parameters:\\
	$\eta_{\nu}=\frac{<m_\nu >}{m_e}$ , $<m_\nu > ~ = ~ \sum^{3}_k~ (U^{(11)}_{ek})^2 ~ e^{i \alpha_k} ~ m_k$	
$	\eta^L_{_N} ~ = ~ \sum^{3}_k~ (U^{(12)}_{ek})^2 ~ 
	~~ e^{i \Phi_k} ~ \frac{m_p}{M_k}$
		with $\alpha_k$, $\Phi_k$ Majorana phases and
$m_p$ ($m_e$) being the proton (electron) mass.
 We see now that we have two types of interference. 
\begin{itemize}
\item The interference between the various left handed contributions.\\
It is thus possible that the light neutrino mass mechanism may be cancelled by the other contributions, so that 
the experiments go below the inverted hierarchy and still do not see the $0\nu$ double beta decay. This, however,
cannot happen in all nuclear systems, since the nuclear matrix elements behave very differently. The light neutrino
operator is long range, but the one resulting from heavy particle exchange is short ranged. 
To be more specific consider { two left handed mechanisms}, one light ($\nu$) and one heavy H. Then
\beq
\frac{m_e/\Omega_{\nu}}{\sqrt{T_{1/2}(0\nu)G_{01}^{0\nu}}}=<m_{\nu}>+\eta_H m_e \frac{\Omega_H}{\Omega_{\nu}}
\eeq
or
\beq
\eta_H=\frac{b-<m_{\nu}>}{m_e r}~~,~~r=\frac{\Omega_H}{\Omega_{\nu}}~~,~~b=\frac{m_e/\Omega_{\nu}}{\sqrt{T_{1/2}(0\nu)G_{01}^{0\nu}}}
\eeq
Now for two targets $A_1$ and $A_2$ we get
$${ <m_{\nu}>=\frac{b_1 r_2-b_2 r_1}{r_2-r_1}}$$
{ More than two targets overdetermine the system and provide tests.}\\
If only the light neutrino mechanism is operative\\
$\Leftrightarrow b_1=b_2=b$ we get the standard expression $ <m_{\nu}>$, which is independent of the target: 
In an analogous fashion  we can write:
\beq
<m_{\nu}> \rightarrow  <m_{\nu}>\left [1+(\eta^L_N/\eta_{\nu}) )(\Omega_N (A) /\Omega_{\nu} (A) \right]
\eeq
Assuming that the cancellation in the target $A_0$ is almost complete (given by c, $c<<1$), we find
\beq
<m_{\nu}> \rightarrow  <m_{\nu}>\left [1-c\frac{\Omega_{\nu}(A_0)}{\Omega_{\nu}(A)}\frac{\Omega_{N}(A)}{\Omega_{N}(A_0)} \right]
\eeq
Even though for the neutrino mass mechanism recent and more reliable nuclear matrix elements have 
appeared \cite{RFSV03}, we
need the matrix elements for the other mechanisms within the same QRPA model. So we are going to use the set 
provided by the earlier paper \cite{PSVF99} (without p-n pairing). The factor inside the square bracket is plotted as a function of A for $A_0=^{76}$Ge and $A_0=^{136}$Xe in 
Figs \ref{fig:frGe} and \ref{fig:frXe}.
 \begin{figure}[h]
\begin{minipage}{18pc}
\includegraphics[width=18pc]{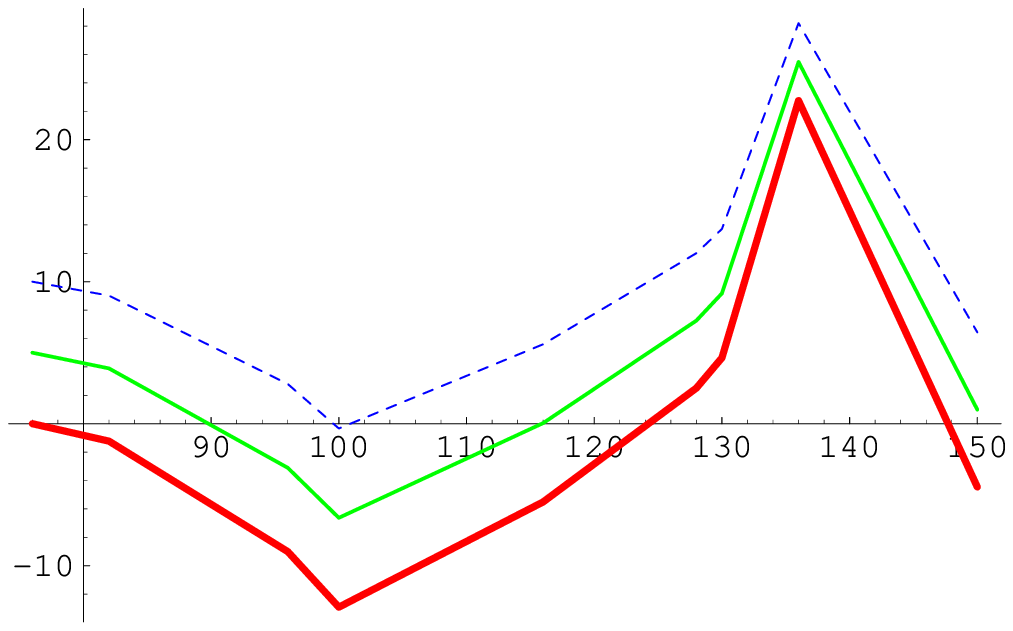}
\caption{\label{fig:frGe} The apparent $<m_{\nu}>$ when there exists 100$\%$ (thick curve), 90$\%$ (fine curve)
and 80$\%$(dotted curve) cancellation 
in the target $A_0=^{76}$Ge.
}
\end{minipage}\hspace{2pc}%
\begin{minipage}{18pc}
\includegraphics[width=18pc]{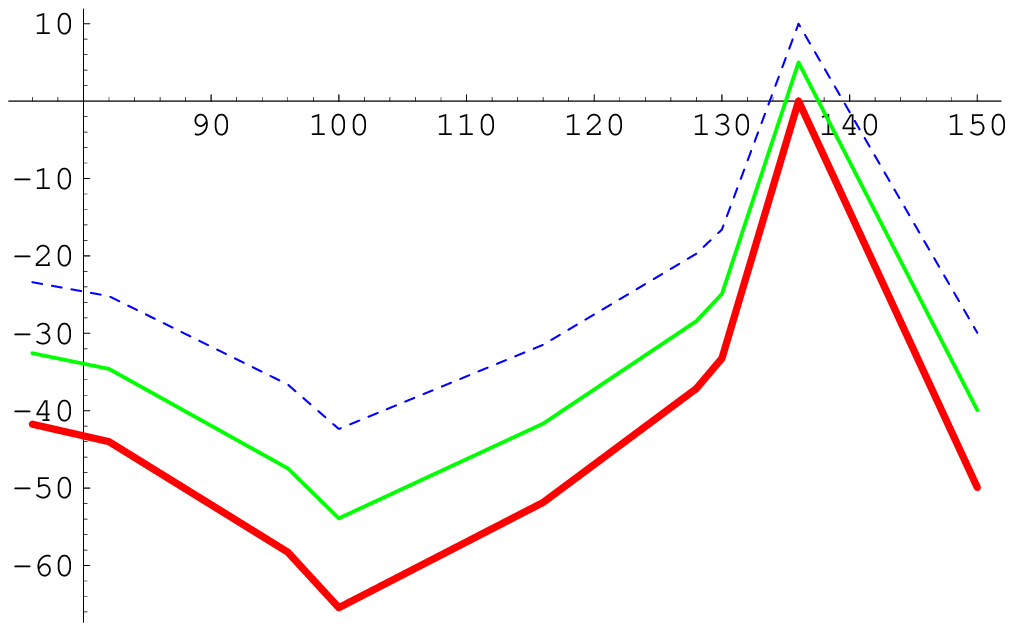}
\caption{\label{fig:frXe} The same as in \ref{fig:frGe} in the case of $A_0=^{136}$Xe.}
\end{minipage} 
\end{figure}

It is clear that such a cancellation cannot occur in all nuclear systems.
\item Interference between left and right handed lepton currents.\\
One can have two amplitudes independent of the neutrino mass if the chirality of the neutrinos is opposite. 
The associated lepton violating parameters are indicated by $\lambda$ and $\eta$ given by:
\beq
{\ \eta} ~ = ~{ \epsilon}~~ \eta_{RL}~~,~~{ \lambda} ~=~ {\kappa}~~
{ \eta_{RL}}
\eeq
\beq{ \eta_{RL}} ~=~
     \sum^{3}_j~ (U^{(21)}_{ej}U^{(11)}_{ej}) e^{i \alpha_j}
\eeq
\beq
      { \kappa}=m^2_L/m^2_R~~,~~{ \epsilon}=tan \zeta.
      \eeq
 $$m_L,m_R= \mbox{ gauge boson masses}~
\zeta= \mbox{ the gauge boson mixing angle}.$$
In the experimental limit on  $^{76}Ge$ after using the nuclear matrix elements of \cite{PSVF99} we
obtain the constraint on the $<m_{\nu},\lambda$ and  $<m_{\nu},\eta$ plane shown in Figs \ref{fig:mlam} 
and \ref{fig:meta} respectively. 
 \begin{figure}[h]
\begin{minipage}{18pc}
 \rotatebox{90}{\hspace{-0.0cm} {$\left <m_{\nu} \right >\rightarrow $ eV}}
\includegraphics[width=18pc]{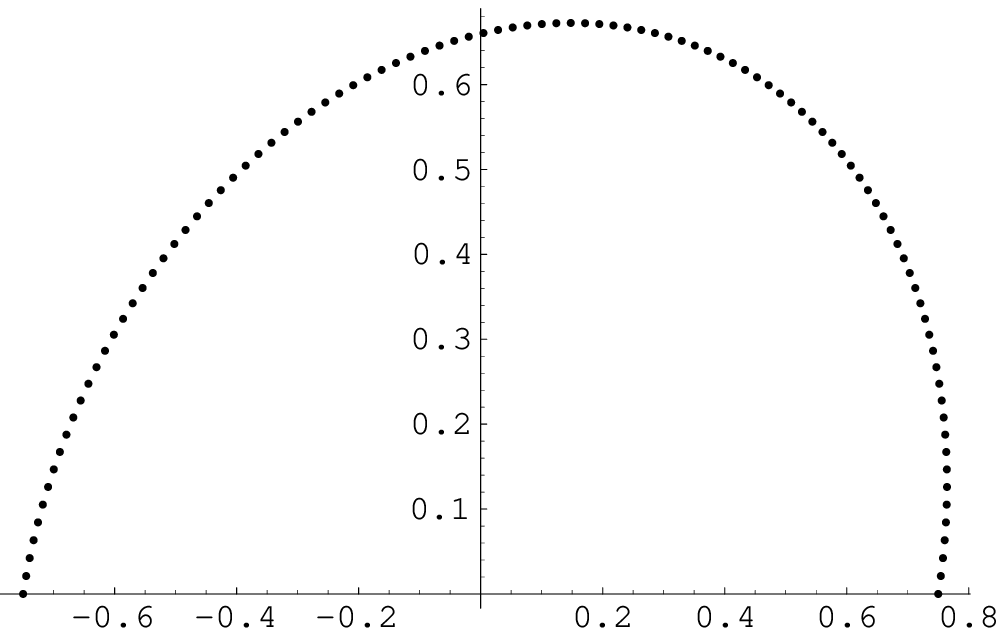}
\hspace*{5.0cm}$\lambda \rightarrow10^{-6}$
\caption{\label{fig:mlam} The constraints on  $<m_{\nu}>$ and $\lambda$ for the target $^{76}Ge$ assuming that they are relatively real.}
\end{minipage}\hspace{2pc}%
\begin{minipage}{18pc}
 \rotatebox{90}{\hspace{-0.0cm} {$\left <m_{\nu} \right >\rightarrow $ eV}}
\includegraphics[width=18pc]{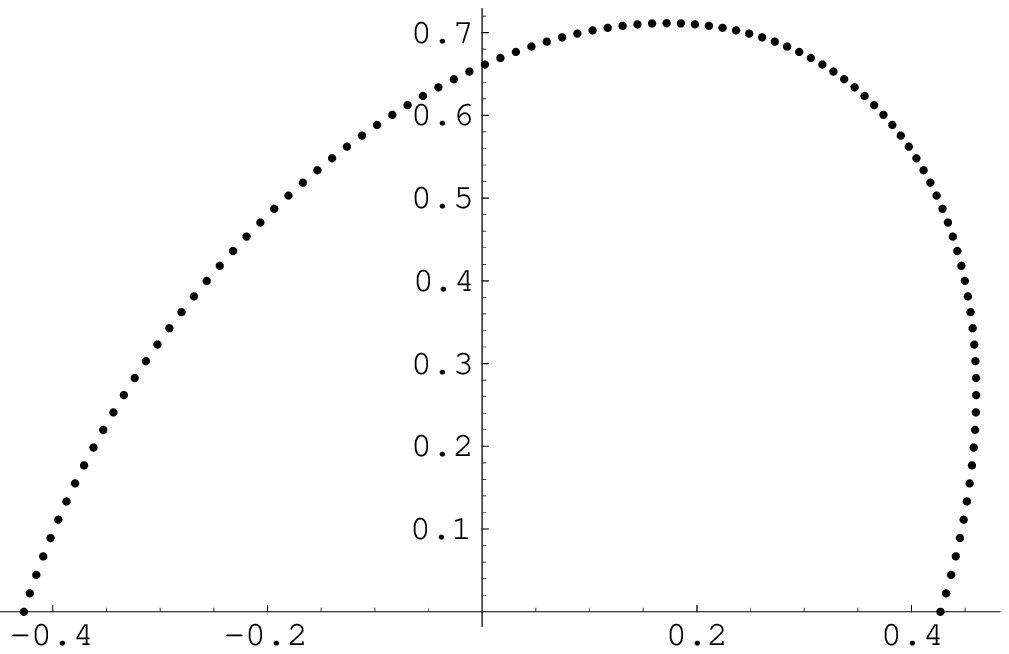}
\hspace*{5.0cm}$\eta \rightarrow10^{-8}$
\caption{\label{fig:meta} The constraints on  $<m_{\nu}>$ and $\eta$ for the target $^{76}Ge$ assuming that they are relatively real.}
\end{minipage} 
\end{figure}
\end{itemize}
\section{Conclusions}
In this brief review we considered some aspects of neutrino physics, which in the frontiers of research
after the discovery of neutrino oscillations.

We first considered neutrinos as probes to detect and study supernova explosions.
 We have  seen that it is quite simple to detect typical supernova
neutrinos in our galaxy, provided that such a supernova explosion takes place (one explosion every 30 years is estimated \cite{SOLBERG}). The idea is to employ a small size spherical TPC detector filled with a high
pressure noble gas and measure nuclear recoils. An enhancement of the neutral current component is achieved via the coherent
effect of all neutrons in the target. Thus employing, e.g., Xe at $10$ Atm, with a feasible threshold energy
of about $100$ eV in the detection of the recoiling nuclei,
 one expects between $400$ and $1900$ events, depending on the quenching factor and the nuclear form factor.
We believe that networks of such dedicated detectors, made out of simple, robust and cheap technology,
 can be simply managed by an international scientific consortium and operated by students. This network
 comprises a system, which can be maintained
for several decades (or even centuries). This is   is a key point towards being able to observe
 few galactic supernova explosions.
 
 We then examined processes which depend on the mixing and mass of the neutrinos. We have shown that
 there are many advantages in using very low energy neutrinos in the few keV and detecting them via 
measuring electron recoils using low threshold gaseous spherical TPC detectors. Then one can look
for oscillations induced by the small mixing angle $\theta_{13}$ and the the small oscillation length,  
associated with the large $\Delta m^2_{23}$, of tens of meters. 
Then the full oscillation takes place inside the detector. With the realistic goal of measuring the distance
up to $1\%$ we hope to measure or put stringent limits on the small mixing angle $\theta_{13}$. At the same
time such an experiment will put a limit on the neutrino magnetic moment at the level of $10^{-12}\mu_b$. It
can also measure the Weinberg angle down to essentially zero momentum transfer.

We have also seen that neutrinoless double beta decay is the only process to decide whether
neutrinos are Dirac or Majorana particles and the best process to settle the question of the
absolute scale of neutrino mass. In order to be able to do so reliable nuclear matrix elements
must be available. Furthermore, even though the neutrino mass mechanism is the most popular 
scenario in
this post neutrino oscillation period, other mechanisms may contribute or even dominate
 this process. Interference between such mechanisms may invalidate any conclusions we draw
about the neutrino mass scale. This indeed maybe a problem, if one analyzes  the data of only one
target. We have seen, however, that such ambiguities may be resolved, if one analyzes data 
from many targets or experiments with different signatures, e.g. tracking versus deposited 
energy.
 
 \ack The author is indebted to the organizers of the Third Symposium on Large TPC's for Low Energy Rare Event Detection, Paris, Dec. 11-12, 2006
  and NUMMY07 ENTApP Network Meeting, Durham,  Jan. 10-12, 2007 for their support in attending the meetings
 where this work was presented.

\section{References}


\end{document}